\documentclass[iop]{emulateapj}
\usepackage{natbib,amsbsy}
\usepackage{amsmath}
\usepackage{graphicx}
\usepackage{subfigure}

\slugcomment{Draft \today}
\shorttitle{Connecting the Dots}
\shortauthors{Ward et al.}

\begin{document}

\title{Connecting the Dots: Analyzing Synthetic Observations of Star-Forming Clumps in Molecular Clouds}

\author{Rachel L. Ward, James Wadsley, Alison Sills, and Nicolas Petitclerc}
\affil{Department of Physics and Astronomy, McMaster University, Hamilton, ON, L8S 4M1, Canada}

\begin{abstract}

In this paper, we investigate the extent to which observations of molecular clouds can correctly identify and measure star-forming clumps. We produced a synthetic column density map and a synthetic spectral-line data cube from the simulated collapse of a 5000 M$_{\odot}$ molecular cloud.  By correlating the clumps found in the simulation to those found in the synthetic observations, clump masses derived from spectral-line data cubes were found to be quite close to the true physical properties of the clumps.  We also find that the `observed' clump mass function derived from the column density map is shifted by a factor of $\sim$ 3 higher than the true clump mass function, due to projection of low-density material along the line of sight.  \citet{alves07} first proposed that a shift of a clump mass function to higher masses by a factor of 3 can be attributed to a star formation efficiency of 30\%.  Our results indicate that this finding may instead be due to an overestimate of clump masses determined from column density observations.
\end{abstract}


\section{Introduction}\label{sec:Intro}

It is believed that most, if not all, stars form in a cluster environment \citep{ladalada03} within highly-compressed cold clouds of dense molecular hydrogen gas and dust.  When observing star-forming regions, we look for evidence of small, compact regions called ``clumps'' -- localized dense regions in giant molecular clouds. These clumps are sites of star formation, and many will continue to collapse to form stars. Understanding star formation regions requires a detailed understanding of the properties and evolution of clumps.  

Recent studies \citep{reid10, CR10, shetty10, pineda09, smith08} have shown that limitations of current observing techniques make it difficult to correctly identify and measure properties of these clumps and cores that reflect the true nature of the star-forming regions.  Observations of star-forming regions in the night sky lack one fundamental component: the third spatial dimension.  It is important to understand how the projection of a three-dimensional structure can introduce bias into the analysis of observations and affect subsequent conclusions.  Issues which arise from projection effects, assumptions regarding cloud properties, and methods used to extract and analyse data can impact the outcome and interpretation of results.  While the Herschel and ALMA observations will disentangle some of these issues, the most self-consistent way to understand the observational biases is by using large-scale simulations to model the collapse of a molecular cloud.

In recent years, many groups have been interested in studying star formation using simulations \citep[e.g.][]{federrath10, offner09, bate09, TP04}.  These types of simulations are able to provide a complete picture of the collapse of the molecular cloud, revealing the various effects of turbulence and gravity.  Simulations of star-forming clouds can also provide insight into physical parameters without confusion from projection effects.  

Using the Smoothed Particle Hydrodynamics (SPH) code, \textsc{Gasoline}, \citet{nic} modelled the collapse of a large 5000 M$_{\odot}$ molecular cloud.  In order to make a direct comparison with observations, we produced synthetic observations from his results.  We compare the clumps found using a common observer's tool -- the publicly-available clump-finding algorithm, Clumpfind \citep{clumpfind} -- on the three-dimensional simulation to those found from the simulated observations to show the discrepancy between the properties of `observed' clumps and `actual' clumps.  

Clumpfind is one of many algorithms currently used for source extraction.  Several new techniques have been developed in recent years, including \textit{getsources} \citep{sasha, sasha2012} and dendrogram methods \citep[e.g.][]{erik2008, shetty10, kauf}.  An exploration of these alternate methods of clump identification is beyond the scope of this paper and has been left for future work.

Studying the earliest stages of star formation can provide clues as to the origin and universality of the stellar initial mass function (IMF), the distribution of stellar masses at the time of their birth.  Since it has been shown that stars form from dense cores \citep[e.g.][]{kirk06,motte98,doug2000}, many believe that the form of the stellar IMF is fixed by the mass distribution of the clumps and cores.  \citet{motte98} extracted a clump mass function (CMF) from observations of the $\rho$ Ophiuchi main cloud and were the first to note the similarity to the stellar IMF.  In the seminal work by \citet{alves07}, it was suggested that the stellar IMF was a direct result of the CMF and that there exists a one-to-one correspondence between the two.  This assumption has largely been adopted as the explanation for the resemblance of the two mass distributions.  However, \citet{reid10} argue that the shape itself is a result of the central limit theorem and that this similarity is not physically significant.  \citet{alves07} also argue that the two mass functions differ only by a uniform star formation efficiency factor of approximately 30\%.  This is implied by a shift of the CMF to higher masses by a factor of 3.  Although many are now questioning whether the IMF is universal across all star-forming regions and if the CMF has a direct one-to-one correspondence with the IMF \citep[e.g.][]{ smith08,swiftwilliams,reid10,anath2011,Michel11}, no alternative explanation has been offered for the factor of 3 difference and many authors continue to cite this factor of 3 as a core-to-star conversion factor \citep[e.g.][]{enoch08,offner09,parm11}.  

In this paper, we will explore these issues using our simulated observations.  In section 2, we provide details of the \citet{nic} simulation and in section 3, we explain the methods used to create our synthetic observations from his results.  In section 4, we present our results followed by a discussion of these results in section 5.

\section{The Simulation}\label{sec:Sim}

Our group has previously simulated star formation in molecular clouds \citep{nic} using the smoothed particle hydrodynamics (SPH) code, \textsc{Gasoline} \citep{gasoline}. We modelled the collapse of an initially spherical 5000 M$_{\odot}$ giant molecular cloud, initially 8 pc across, using approximately 36 million particles.  The key specifications of the simulation are listed in Table~\ref{table:initcondit}.  We use an equation of state from \citet{BB05} to model the thermal behaviour of the gas.  We have chosen this equation of state, as it mimics the behaviour of temperature and density seen in spherically-symmetric collapses of molecular clouds using radiative transfer \citep{BBB03}. The value of 10$^{-13}$ g cm$^{-3}$ in this equation of state is taken to be the opacity limit for fragmentation \citep{BBB03,BB05}.  Since our maximum density of 10$^{-15}$ g cm$^{-3}$ is well below this density limit of 10$^{-13}$ g cm$^{-3}$, the entire simulation is isothermal at 10K and the gas is optically thin.  In this paper, we present the results at 1.3 million years after the simulation began -- approximately 70\% of the initial free-fall time. 

\begin{table}
	\caption{Initial conditions of the SPH simulation \\ 
	of a giant molecular cloud}
	\centering
	\begin{tabular}{c c c c}
	\\
	\hline\hline   
	Mass	&												5000 M$_{\odot}$ \\
	Particles	&										36 088 472 \\
	Diameter	&										8 pc  \\
	Initial uniform density & 		300 cm$^{-3}$ \\
	Initial Jeans mass & 			2.8 M$_{\odot}$	\\
	Particle mass	&								1.3855 $\times$ 10$^{-4}$ M$_{\odot}$ \\
	Free-fall time 	&							1.9 Myr \\
	Temperature 	&								10 K	\\
	Minimum Jeans mass	&					0.01 M$_{\odot}$ \\
	Minimum smoothing length	&		50 AU \\
	Minimum spatial resolution	& 100 AU \\ [1ex]
	\hline
  \end{tabular}
  \label{table:initcondit}
\end{table}

These simulations had open boundary conditions, allowing us to simulate an entire molecular cloud undergoing gravitational collapse, rather than just a small portion within it.  Other simulations have tended to focus solely on sub-regions within the clouds with masses only reaching as high as 500 solar masses \citep[e.g.][]{bate09,smith08,TP04}.  However, these sizes are typical for a single clump seen in observations \citep{BT07}.  Thus, in order to more accurately represent what is seen in observations, we use simulations of star formation that cover much larger scales.  

\section{Simulated Observations}\label{sec:SimObs}

In order to compare the results of the simulation with observations, we have created synthetic observations of a three-dimensional (position-position-velocity or `PPV') spectral-line data cube and a two-dimensional (position-position or `PP') column density map.  These data are a more realistic representation of observations, as observers typically do not know the physical position of an object along the line of sight.  Our simulation provides the third spatial dimension necessary for comparison to the synthetic observations (position-position-position, or `PPP'). 

\subsection{PPP}

To compare real objects found in our simulations to objects seen in our simulated observations, we need to characterize the simulation in a way that is complementary to the observations.  Therefore, we created a 3D smoothed density map of the gas in the simulation by interpolating the SPH particles to a 500 x 500 x 500 grid. The full spatial extent of the simulation is 3 $\times 10^6$ AU on a side so each pixel corresponds to 6000 AU.  Any physical quantity associated with the gas can be estimated by interpolating between the function values at the known positions of the particles.  A kernel function \citep{ML85} converts the particle information into a smoothed estimate.  The weighting determines what fraction of the particle mass will contribute to each grid cell.  The sum of the kernel weights is normalized to ensure the total mass of the particle is counted only once.  If the size of the particle carries it off the edge of the grid, it is assigned to the edge of the grid and if there are no grid lines intersecting the smoothing sphere, the particle weight will be assigned to the nearest grid cell.  The cube can be rotated and ``observed'' along any line of sight.

\subsection{PP}

\begin{figure} 
\begin{center}
\includegraphics[scale=0.4, angle=0]{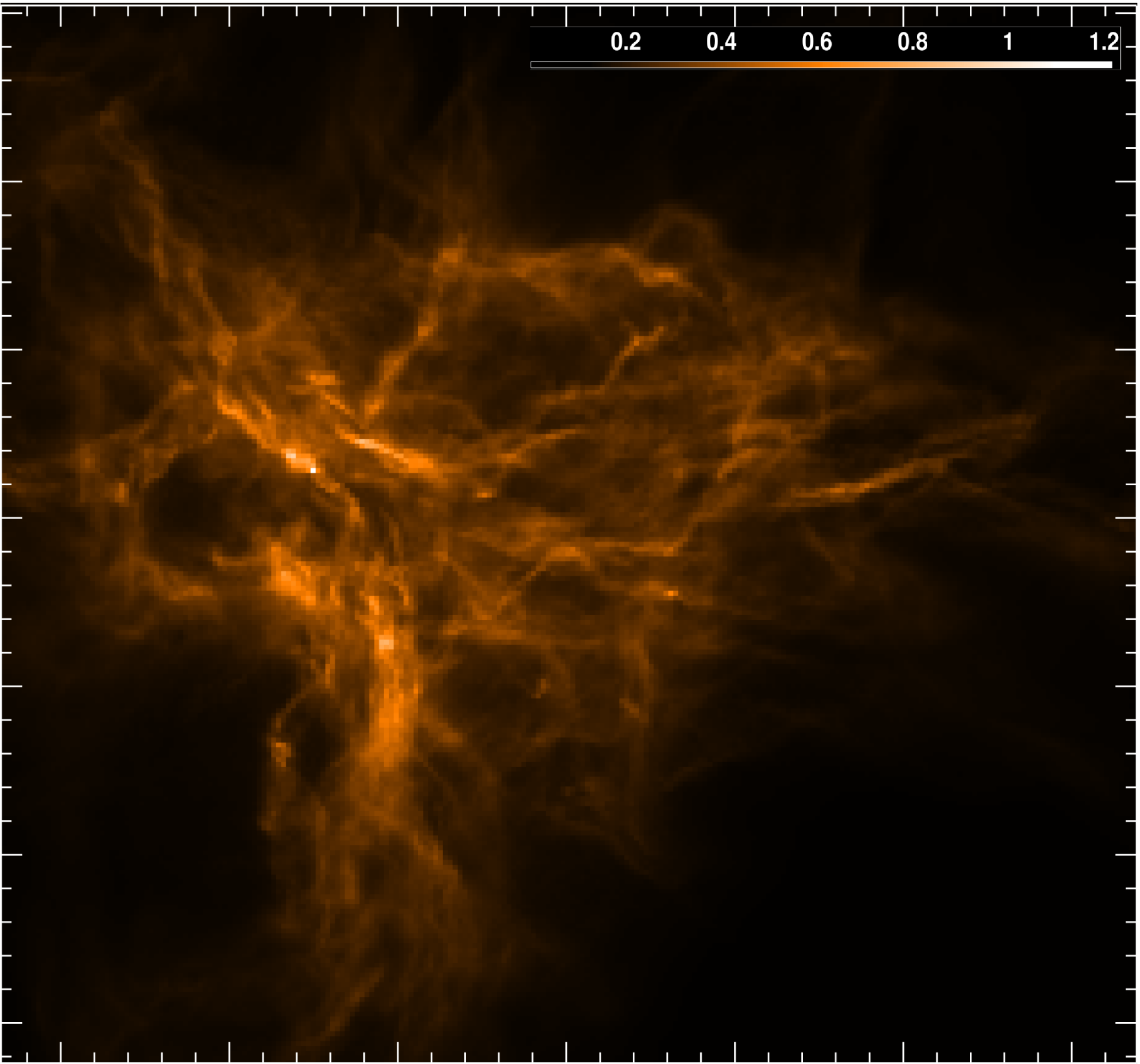}
\caption{\label{fg:XYcol6000fullcolour}
 	Two-dimensional column density map of the simulated molecular cloud. The resolution of the map is 6000 AU$/$pixel.  The total size of this box is 1.25 $\times$ 10$^6$ AU by 1.25 $\times$ 10$^6$ AU.  Note the filamentary structure comparable to that seen in observations.  The column density is in units of solar mass per pixel.
  }
\end{center}
\end{figure}

Star-forming regions are often observed in dust continuum observations, which provides a measure of the column density of gas along the line of sight. Our two-dimensional (PP) column density map was produced from the simulation by integrating the density along the line of sight using the \emph{collapse} task from the \textsc{kappa} software package, a part of the Starlink software environment\footnote{http://starlink.jach.hawaii.edu}.  We arbitrarily chose the z axis of the simulation as our line of sight. The two-dimensional column density map produced can be seen in Figure \ref{fg:XYcol6000fullcolour}. We assume that the column density of material is directly analogous to the observed flux in sub-mm or IR observations since our gas is optically thin throughout the simulation time. Any heating which occurs in these relatively unevolved cores due to accretion is immediately radiated away.

The spatial resolution is the same as the PPP cube:  6000 AU/pixel and an image size of 500 x 500 pixels.   The resolution was chosen to be comparable to that of an observation of the active Orion star-forming region using the SCUBA-2 instrument on the James Clerk Maxwell Telescope (JCMT). For the synthetic column density map with this resolution, the densest gas would correspond to a column density of approximately 1 M$_{\odot}/$pixel or 10$^{23}$ cm$^{-2}$.  These peak values are comparable to those seen in real observations, as starless cores have been observed to have typical column densities $<$ 10$^{23}$ cm$^{-2}$ \citep{caselli02a}.  

\subsection{PPV}

Submillimetre observations using spectral imaging systems, such as the HARP and ACSIS instruments \citep{buckle} on the JCMT, allow astronomers to record a third dimension in the star-forming region.  This third dimension is not a spatial dimension, but rather the velocity of the gas along the line of sight.  This observational technique produces images at various wavelengths to form a spectral-line data cube.   

To create the synthetic  spectral-line data cube, we use the velocities of the particles to determine the motion of the gas along the line of sight (which we again take to be the z axis).  For each particle, a contribution was added to the appropriate velocity bins (or channels) for all lines of sight the particle intersects.  This produces infinitely sharp `spectral lines'.  For a given velocity channel, each contributing particle mass in that bin is divided up into grid cells by their specific weights.  Combining each frame of column density per velocity channel results in a cube with two spatial dimensions and one velocity dimension. 

Our aim is to model molecular line emission along the line of sight; therefore, we require a molecular tracer to simulate the spectral-line emission.  We have chosen the NH$_3$ (1,1) hyperfine transition as our molecular line tracer.  NH$_3$ is an excellent tracer of dense cores, as it is present at very high densities \citep{caselli02a,caselli02b,BT07}.  Our PPV cube is highly resolved with 148 velocity bins with a resolution of 0.04 km s$^{-1}$ per bin, which corresponds to a 3.050 kHz channel separation for the NH$_3$ spectrum.  This was chosen to approximately match that in the NH$_3$ observations of a dense core used by \citet{pineda10}.  

Since actual molecular lines have a finite width, we must introduce a method to broaden our lines.  We have chosen to use thermal broadening.  To determine the width of our lines, we used the temperature of the isothermal gas in our simulation (10 K) and the molecular mass of NH$_3$.  Using our velocity bin width and corresponding frequency channel separation, as well as the rest frequency of NH$_3$, we determined the width to be 7.6 kHz or 0.1 km s$^{-1}$.  This width, together with the line-of-sight velocities (and corresponding frequencies) of the particles, allow us to broaden each of the spectral lines in our PPV cube.

We produced a PPV cube where only the gas emission traced by NH$_3$ is included.  The NH$_3$ transition is excited in cloud cores at a critical excitation density of 1800 cm$^{-3}$ \citep{tielens} and NH$_3$ is expected to deplete only at very high densities $\geq$ 10$^6$ cm$^{-3}$ \citep{BL97}.  NH$_3$ is also abundant in dense cores and its continued presence at the high densities of core interiors makes NH$_3$ is one of the best tracers of the physical, chemical, and kinematic properties of dense cores \citep{pagani}. We used a threshold of 1800 cm$^{-3}$ and did not include the contribution of  any gas below this density when making this PPV cube.

\section{Clump-finding and Correlations}\label{sec:clfind}
\subsection{Studying Clumps in the Third Dimension}

With our simulation of molecular cloud collapse, we have `observed' the clumps formed within the cloud using the clump-finding algorithm, Clumpfind \citep{clumpfind}, which we will refer to as Clumpfind3D when applied to 3D data. The Clumpfind algorithm requires that we define a value for $\sigma$, which in observations is the noise level.  The algorithm then determines intensity contours in multiples of $\sigma$ for the entire dataset. The user specifies step increments and the minimum contour level in multiples of $\sigma$. The algorithm moves through the dataset in steps of the specified increment and assigns pixels to clumps based on the landscape of the local contour levels.  The algorithm stops assigning pixels when it reaches the specified lowest contour level. 

The largest single value in a pixel for the PPP data cube was 0.8 M$_{\odot}$.  To obtain an estimate of the Clumpfind3D threshold for the PPP cube, we divided this maximum pixel value by a typical dynamic range.  The dynamic range is defined for observations as the ratio between the brightest intensity and the background noise.  As an example, the dynamic range on the Herschel instruments, PACS and SPIRE, are $\sim$1000 and $\sim$200, respectively \citep{motte10}.  A conservative estimate of 160 was used as the dynamic range for the PPP cube, resulting in a $\sigma$ value of approximately 0.005 M$_{\odot}$/pixel for a resolution of 6000 AU/pixel.  We used this dynamical range to set the Clumpfind threshold rather than adding noise to our simulations because the choices for the noise properties would be arbitrary. We have also shown \citep{reid10} that the derived properties of clumps are basically independent of noise levels, when the dynamic range of the observations are comparable to that of modern instruments.

Setting the lowest density contour level in Clumpfind3D to 3$\sigma$ and the contour increment to 2$\sigma$ results in the identification of 499 clumps in the PPP data cube. These clumps contain approximately 430 M$_{\odot}$ of material or 9\% of the total mass in the simulation.  

\textbf{}

\subsection{Clump Correlations, Round One: PP vs. PPP}

We ran the two-dimensional implementation of Clumpfind (which we will refer to as Clumpfind2D in this paper) on the synthetic column density map. All clumps were identified using the \emph{findclumps} task from the source extraction software package \textsc{cupid}, which is a part of the Starlink software collection. We used a variety of different parameters to determine their effect on the number of clumps identified by the algorithm.  Our results are shown in Table~\ref{table:clfind2dparam}, where `Low' is the lowest contour level and `Step' is the contour interval.  Although changing the parameters can cause the number of identified clumps to vary by a factor of up to 2, the amount of mass in clumps is independent of the parameters chosen.  The mass in clumps makes up $>$ 90 \% of the mass of the entire cloud.  Clumpfind2D decomposes the cloud into clumps by assigning almost all material to clumps.  For this reason, many of these clumps may not necessarily connect to any physical structures within the cloud. The inclusion of all emission along the line of sight, even low-density gas unassociated with the dense region, could affect the determination of clump properties.

\begin{table}
	\caption{Clumpfind2d Parameter Sensitivity}
	\centering
	\begin{tabular}{c c c c c c c}
	\\
	$\sigma$ & Low & Step & \# of clumps & Mass in clumps  & Percentage of\\
	(M$_{\odot}$/px) & ($\sigma$) & ($\sigma$) & & (M$_{\odot}$) & Total Mass\\
	\hline\hline   
	0.005  & 3$\sigma$ & 2$\sigma$ & 886 & 4937 &99\% \\
	0.006  & 3$\sigma$ & 2$\sigma$ & 558 & 4777 &96\% \\
	0.007  & 3$\sigma$ & 2$\sigma$ & 532 & 4723 &94\% \\
	0.006  & 5$\sigma$ & 3$\sigma$ & 455 & 4558 &91\% \\
	0.005  & 5$\sigma$ & 3$\sigma$ & 503 & 4648 &93\% \\ 
	\hline
  \end{tabular}
  \label{table:clfind2dparam}
\end{table}

It has previously been shown that clump properties determined from Clumpfind2D can be unreliable \citep{kain2009}.  Our results confirm that a small change of parameters in a clump-finding algorithm can alter both the number of clumps located and their properties. It is difficult to obtain sets of parameters for the Clumpfind algorithm for the PP and PPP cases that identify the same clumps in both `observations'. Rather than identify clumps independently in both PP and PPP, we have taken the clumps identified in the three-dimensional volume and determined their position and area in the $xy$ plane. We used a pixel mask made from this projection of clumps from the PPP cube and identified them directly onto the PP map. The mass of each PP clump includes all the mass along the lines of sight within the area traced out by the PPP clump. This correlation method ensures that the clumps have an equal projected area and pixels with identical (x,y) coordinates.

\begin{figure}
\begin{center}
\includegraphics[scale=1.0, angle=0]{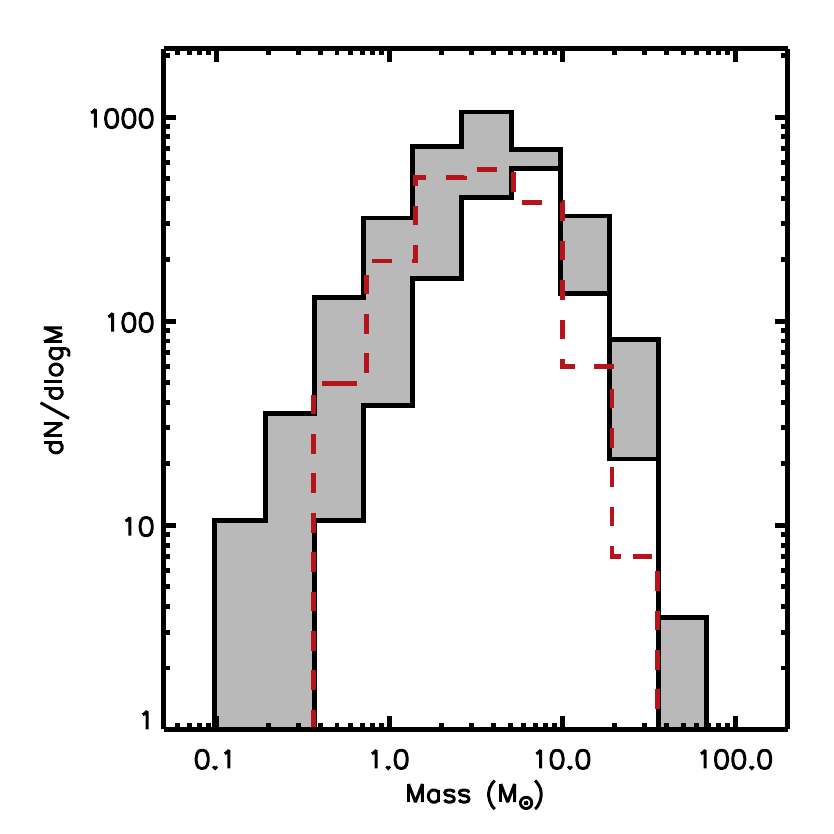}
\caption{\label{fg:cmfPPcompare} 
 The shaded region shows the range of possible mass distributions for clumps identified from the PP column density map using Clumpfind2D for the cases listed in Table 2.  This is compared to a clump mass function of clumps identified from the same PP column density map using our correlation method (dashed line).  
 }
\end{center}
\end{figure}

\begin{figure*} 
\begin{flushright}
\includegraphics[scale=0.9]{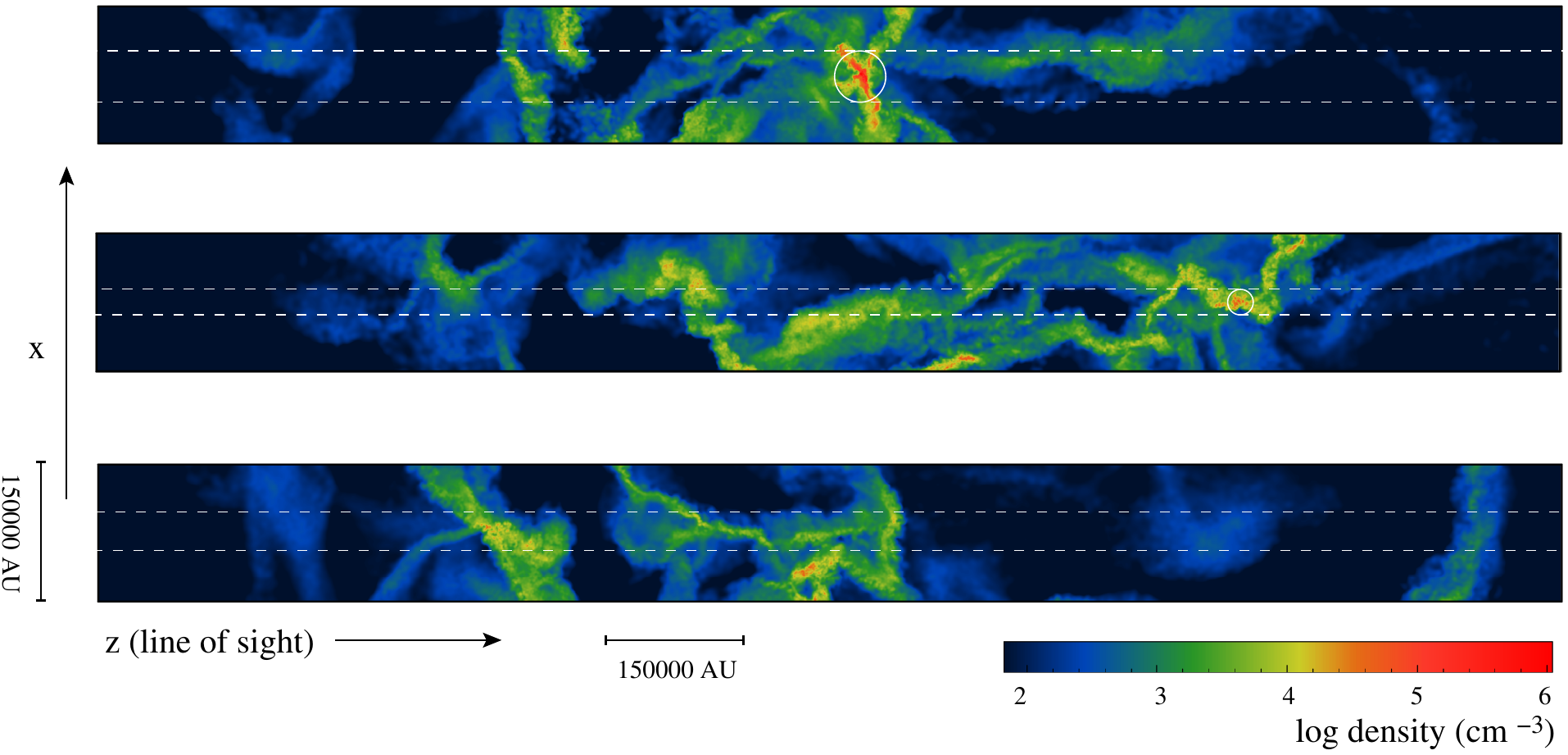}

\end{flushright}
\caption{\label{fg:columns} 
	Cross-section slices of the simulation in the xz-plane.  The colour scale shows density in units of cm$^{-3}$. Large filaments of low density gas are present along the z-axis, chosen to be our line of sight, when observed in the xz-plane. 
The top and middle panels respectively show the largest clump and a typical clump (both circled) identified in the PPP cube plus the foreground and background material which contributes to their PP masses.  These clump masses are shown as circled points in Figure~\ref{fg:PPPvsPPlog}.  The bottom panel shows a region which was identified as a clump in two dimensions using Clumpfind2D, but contains no true clump as identified by Clumpfind3D.  The white dashed lines contain the lines of sight which intersect the circled clump of interest, where the spacing of the white lines is determined by the size of the clump.  			 
	 These slices have a width of 1.5 $\times$ 10$^5$ AU and a length of 1.5 $\times$ 10$^6$ AU, extending fully across the simulation. 
	This visualization was made using SPLASH \citep{price07}.
	}
\end{figure*}

Figure~\ref{fg:cmfPPcompare} compares the CMFs of clumps identified in the PP column density map using the Clumpfind2D method and our correlation method.  The correlation method finds the \emph{projected} masses of the PPP clumps by including all the emission along the line of sight.  The shaded region shows the range of distributions produced by Clumpfind2D for the cases listed in Table~\ref{table:clfind2dparam}.  The range is tightly constrained at the high-mass end and the CMF produced by the correlation method provides a lower limit to the range of possible mass distributions produced by the Clumpfind2D method.  

\subsubsection{Global Properties}

Clumpfind3D identified 499 clumps in the PPP cube and returned the integrated intensity and the peak (x,y,z) positions of each of these clumps. The total mass contained in clumps in the PPP cube is approximately 430 M$_{\odot}$.  Using the correlation method, the total mass contained in clumps from the masked PP map, which has an equal area in the XY-plane to the clumps in the PPP cube, is 1647 M$_{\odot}$.  This is not surprising as large amounts of low density material along the line of sight are now contributing to the mass estimate of the clumps in those columns (see Figure~\ref{fg:columns}).  When the simulation is integrated along this axis to produce a two-dimensional column density map, all of this emission contributes, resulting in an identified object with a mass which is larger than the real star-forming clump. In some cases (e.g. the bottom panel of Figure~\ref{fg:columns}) there is no clump identified in the PPP cube, but the integrated intensity of a significant amount of low-density gas can masquerade as a clump in the projection on the plane of the sky. 

\begin{figure}
\begin{center}
\includegraphics[scale=1.0, angle=0]{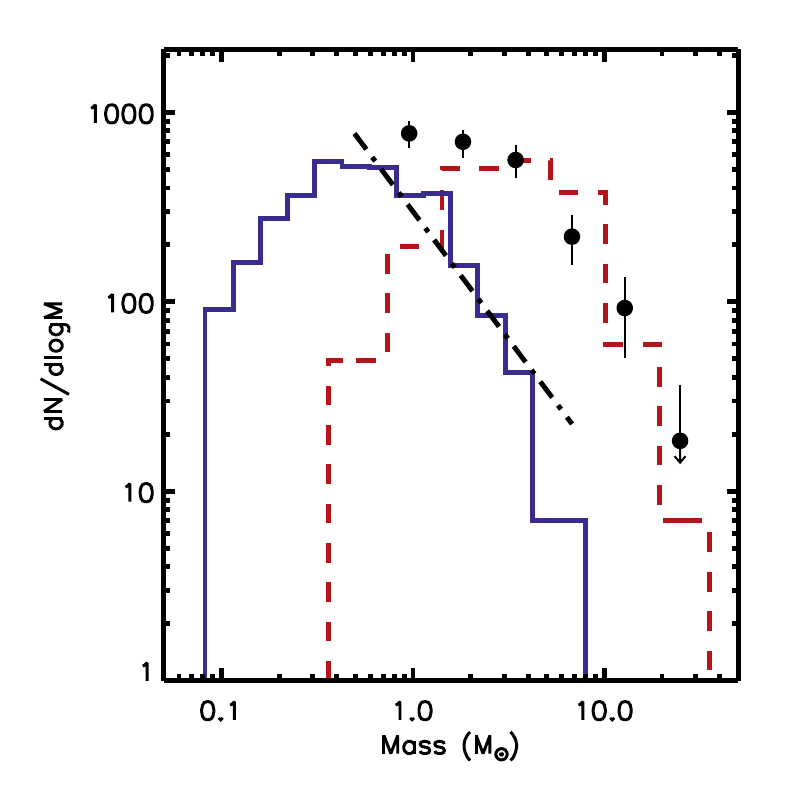}
\caption{\label{fg:cmfPPPvsPP} 
 Clump mass function of the clumps identified in the PPP data cube (solid line).  Also shown is the clump mass function of the same clumps with their masses estimated from the 2D projected PP image using the correlation method (dashed line).  The dash-dotted line is the \citet{salpeter} IMF.  The dense core mass function (circles) obtained by \citet{alves07} from observations of the Pipe Nebula is also shown for comparison.  
 }
\end{center}
\end{figure}

Figure~\ref{fg:cmfPPPvsPP} shows the mass distribution of the 499 clumps for both the PPP case (solid line) and the PP projection (dashed line).  Both clump mass functions closely reflect the power-law trend at high mass seen in the stellar IMF \citep{salpeter}, shown by the dash-dotted line in Figure~\ref{fg:cmfPPPvsPP}.  We have also plotted the \citet{alves07} dense core mass function for the Pipe Nebula, shown in Figure~\ref{fg:cmfPPPvsPP} as circles with error bars. We note that the factor of $\approx$ 3 difference in mass between the PP and PPP clump masses is due entirely to the contribution of low-density gas and multiple cores along the line of sight in the PP case. While our PPP clump mass function is well-fit by a Salpeter mass function, these objects are still clumps, not protostars. The mass difference here cannot be interpreted as a star formation efficiency. 

\subsubsection{Clump-by-Clump Comparison}

\begin{figure}
\begin{center}
\includegraphics[scale=1.0, angle=0]{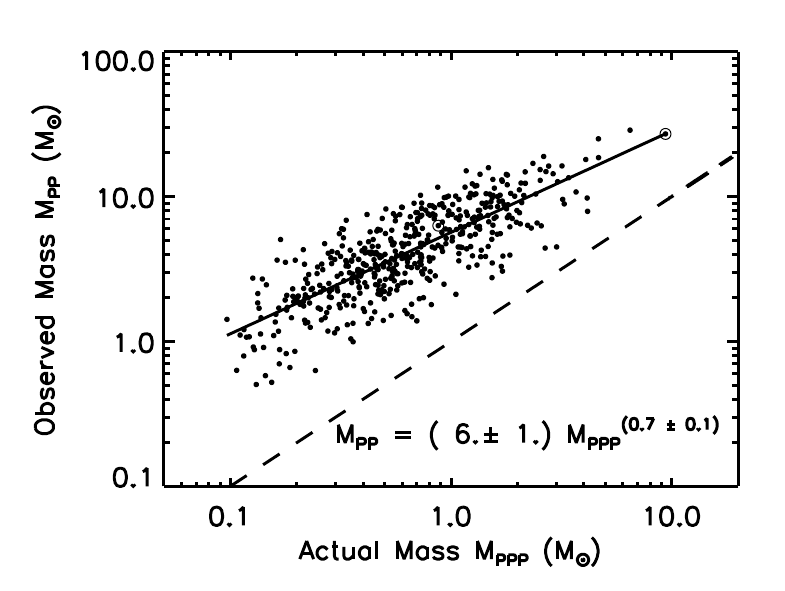}
\caption{\label{fg:PPPvsPPlog} 
Observed mass from the column density map vs. the actual mass from the simulation cube.  A fit to the data is shown as a solid line, and given on the figure. The dashed line shows the expected correlation if the observed mass were not contaminated by foreground and background material.  The circled points are the clumps shown in the top and middle panels of Figure~\ref{fg:columns}.}
\end{center}
\end{figure}

Using the masses obtained for each of the 499 clumps from our correlation method, we have plotted the observed masses of each clump (in PP) against their corresponding actual masses (in PPP) to produce a direct clump-to-clump comparison.  The power-law trend observed between M$_{PPP}$ and M$_{PP}$ in Figure~\ref{fg:PPPvsPPlog} exists due to the correlation between M$_{PPP}$ and the projected area of the PPP clumps.  Though the column of material along the line of sight (as shown in Figure~\ref{fg:columns}) is not correlated with the PPP masses, as can be seen from the scatter in Figure ~\ref{fg:PPPvsPPlog}, when multiplied by the area, a net correlation arises.  

The functional form
\[M_{\text{PP}}= (6\;\pm\;1) \left(M_{\text{PPP}}\right)^{0.7 \pm 0.1}\]
was fit to the data with a reduced chi-squared of 3.5 $\times$ 10$^{-2}$. The slope of this relationship is $\approx$ 2/3, and we expect this is related to the ratio of surface area ($\propto$ R$^2$) to volume ($\propto$ R$^3$) of the clump, since the density is approximately constant. For a given PPP mass between 2M$_{\odot}$ and 10M$_{\odot}$, this function can provide a range of possible PP masses that are approximately 3-5 times greater, which is again comparable to the factor of three often quoted in other works with regard to the CMF \citep{alves07,enoch08,offner09}.  The estimated PP masses can be even more than 3-5 times larger than the actual mass for lower mass objects. 

\subsection{Clump Correlations, Round Two: PPV vs. PPP}

A second method of determining properties of star-forming clumps is to observe them using spectral line data. We follow the same method of identifying clumps in the PPP cube and then studying the properties of the same region in the PPV `observation'. In order to obtain a PPV mask of the desired clumps from the PPP cube, we first obtained the (x,y) pixel coordinates of the PPP clumps to create a 2D mask.  We then determined the mean velocities of the PPP clumps to correlate the clumps in (x,y,z) coordinate space with (x,y,v$_z$) coordinate space. 

The mean velocity and velocity dispersion of each of the PPP clumps were determined using the velocities of the particles in the simulation.  We are only concerned with the mean velocity along the line-of-sight, $\bar{v}_z$, for comparison with the PPV cube.  Once the mean velocity and velocity dispersion of each clump are known, we use these values to estimate a range of velocities, [$\bar{v}_z-\sigma_z, \bar{v}_z+\sigma_z$], to represent the range of velocity channels spanned by a Doppler-broadened line profile in the velocity spectrum.  

For each clump, we specify the velocity bins we would like to extract from the PPV cube. We also specify the minimum volume density of gas that can contribute to the NH$_3$ emission (1800 cm$^{-3}$).  The mass of the PPV clumps is the contribution of all particles in high-density regions moving at line-of-sight velocities between $\bar{v}_z-\sigma_z$ and $\bar{v}_z+\sigma_z$, with the same {xy} area as the PPP clumps. We have tested this method of clump selection against the clumps found by applying Clumpfind3D directly on the PPV cube, and the clump mass functions are very similar. Therefore, we analyze the properties of the clumps found by our correlation method in the PPV cube.

\subsubsection{Global Properties}

\begin{figure} 
\begin{center}
\includegraphics[scale=1.0, angle=0]{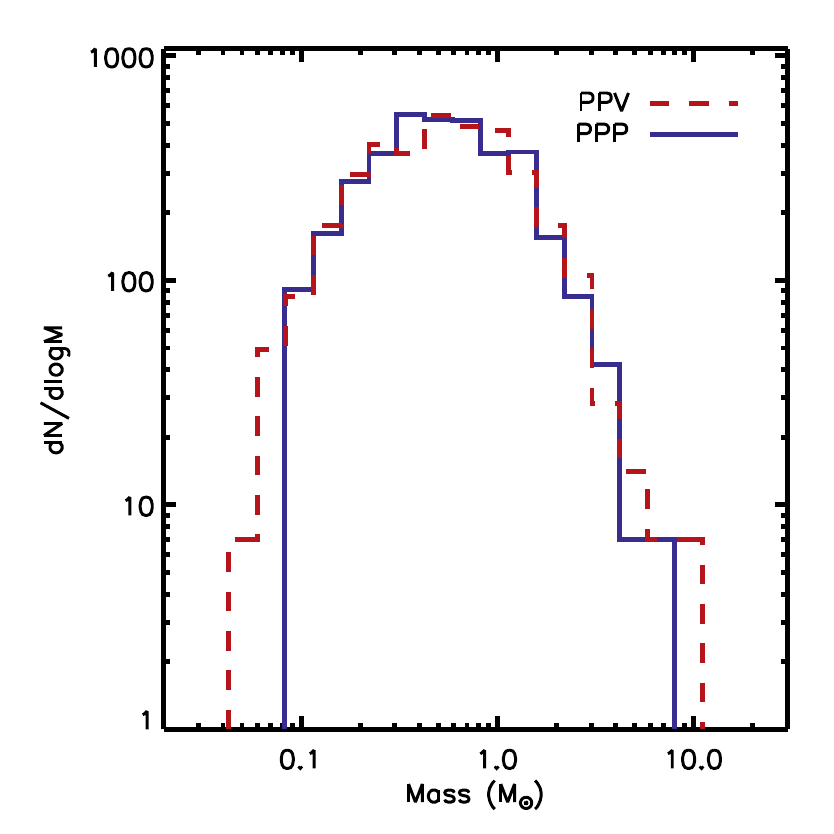}
\caption{\label{fg:cmfPPPvsPPVNH3} 
 Clump mass function of the clumps identified in the PPP data cube (solid line) compared to the clump mass function of the same clumps with their masses estimated from the NH$_3$ (1,1) emission of the (x,y,v$_z$) PPV data cube (dashed line).}
\end{center}
\end{figure}

The total mass contained in clumps in the PPP cube is approximately 430 M$_{\odot}$.  Using the correlation method to locate these clumps in the PPV data cube, we find that the total mass contained in clumps from the masked PPV cube is  439 M$_{\odot}$.  We have produced a clump mass function for the 499 clumps identified in the NH$_3$ PPV data cube, shown in Figure~\ref{fg:cmfPPPvsPPVNH3}.

The PPV clump mass function is extremely close to the true PPP clump mass function.  We performed a two-sided Kolmogorov-Smirnov Test (KS-Test) to compare the PPV and PPP masses.  There was an 81\% probability that the PPV and PPP masses were drawn from the same distribution.  Most of the extended emission, even that  moving at a similar line-of-sight velocity as the clump, does not contribute significantly in this case.  Based on these results, it appears that clump properties derived from PPV spectral-line data cubes are quite representative of the true physical properties of the clumps.  

\subsubsection{Clump-by-Clump Comparison}

\begin{figure}
\begin{center}
\includegraphics[scale=0.8, angle=0]{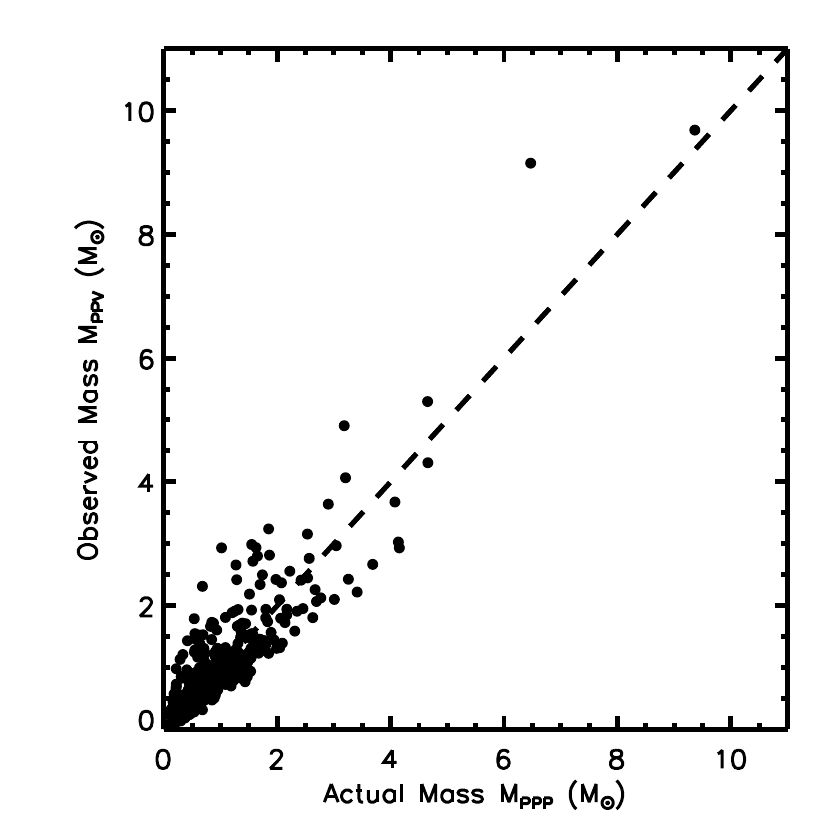}
\caption{\label{fg:PPPvsPPVcompare} 
 Observed mass (PPV) vs. the actual mass (PPP) of clumps. The PPV masses are obtained from only the highest density gas traced by NH$_3$.  The dashed line is the one-to-one line. }
\end{center}
\end{figure}

The comparison of actual and observed masses using the PPV spectral cube are shown in Figure~\ref{fg:PPPvsPPVcompare}. Using the masses obtained for each of the 499 clumps from the correlation method, the observed PPV masses are, on average, the same as the actual PPP masses.   We conclude that using spectral line data is the best method for accurate estimation of clump properties.   

\section{Discussion}\label{sec:discuss}

We have `observed' a SPH simulation of the collapse of a spherical molecular cloud, using the techniques and tools common to real observations of star-forming regions.  
We determine the masses of three-dimensional star-forming clumps as they would be observed in a two-dimensional column density map and in a PPV spectral-line data cube.  We found that projection has the greatest effect on estimates derived from PP column density maps.  We have seen that there are some objects identified in column density maps as clumps, which are not clumps at all. Instead, they are extended regions of low density gas which, when projected onto a two-dimensional sky, appear as a single object.  We found that effects of projection can result in overestimates of clump masses derived from two-dimensional column density maps.  Our results show that the observed shift of the CMF from the stellar IMF to higher masses \citep{alves07} is not the result of a star formation efficiency factor.

By correlating the clumps found in PPP to those found in PPV, we find that the properties of objects derived from PPV spectral-line data cubes were more representative of the true physical properties of the clumps than those obtained from the PP column density map. We conclude that clump properties are best derived from molecular-line data rather than continuum data, in order to minimize projection effects and maximize the identification of truly bound structures in observations. 

In addition to the work presented here, we investigated various modifications of our simulation with different evolutionary stages, different orientations, and for different initial collapse conditions.  Recent studies \citep{arzou,schmalzl,sasha} have shown that filaments are very common in molecular clouds and may have a significant role in star formation.  To learn more about the importance of initial geometry, additional simulations were run with different large scale turbulent velocity modes included in the initial conditions.  When we study the effect of the initial turbulent velocity spectrum on the formation of dense cores, the anisotropic collapse cases all found significantly more dense clumps and more mass contained in clumps than that seen for the spherical collapse case.  This result implies that dense cores in the cloud are more inclined to form in filamentary-like structure, which is in agreement with conclusions drawn from recent observations of star-forming regions \citep{arzou,schmalzl,sasha}.  However, we found that our results regarding the relationship between the `observed' masses and `actual' masses are unchanged, regardless of the evolutionary stage, the orientation, and the initial collapse condition.  

The next step in this work is to look at an even larger (and more realistic) simulation. We are currently running a simulation of a 50000 M$_{\odot}$ molecular cloud to track the evolution of the cores from the initial collapse of the cloud through to the pre-main sequence evolutionary phase.  As the cores evolve, we will be able to see how often they will merge and collapse into bound objects and how often they disperse to determine a statistical likelihood that an observed core will eventually become a star.  The study of the core evolution will provide insight into the formation mechanisms of multiple star systems, such as binaries, and will also give an indication of the multiplicity of these objects in observations; for example, how often a single, identified core is, in reality, two or more widely separated cores which have been confused in projection.

We also intend to run a simulation incorporating the effect of radiative feedback from the protostars using a radiative transfer code recently implemented in \textsc{Gasoline} \citep{rogers}.  Since we are interested in studying the later stages of pre-main sequence stellar evolution when the cores become associated with a compact source of luminosity and are emitting protostellar radiation, we need to understand the effects of localized heating due to the radiative feedback from newly-formed stars and determine the significance of the addition of radiative transfer under the conditions of our simulation. 

\section*{Acknowledgments}

We would like to thank SHARCNET (Shared Hierarchical Academic Research Computing Network) 
and Compute/Calcul Canada, which provided dedicated resources to run these simulations.  
We would also like to thank Christine Wilson and Michael Reid for their helpful conversations. 
This work was supported by NSERC.  JW acknowledges support from the Ontario Early Researcher Award (ERA).
\\
\\
\\
\\
\\
\bibliographystyle{apj}

\end{document}